Jaakko Mäkinen

# The permanent tide and the International Height Reference Frame IHRF

*Finnish Geospatial Research Institute (FGI), National Land Survey of Finland, Geodeetinrinne 2, FI-02430 Masala, Finland*

*Corresponding author Jaakko Mäkinen*
e-mail Jaakko.Makinen@nls.fi, ORCID 0000-0002-7274-6376

## Abstract

The International Height Reference System (IHRS), adopted by IAG (International Association of Geodesy) in its Resolution No. 1 at the XXVI General Assembly of the International Union of Geodesy and Geophysics (IUGG) in Prague in 2015, contains two novelties. Firstly, the mean-tide concept is adopted for handling the permanent tide. While many national height systems continue to apply the mean-tide concept, this was the first time that the IAG officially introduced it for a potential field quantity. Secondly, the reference level of the height system is defined by the equipotential surface where the geopotential has a conventional value $W_0$=62636853.4 m$^2$ s$^{-2}$. This value was first determined empirically to provide a good approximation to the global mean sea level and then adopted as a reference value by convention. I analyse the tidal aspects of the reference level based on $W_0$. By definition $W_0$ is independent of the tidal concept that was adopted for the equipotential surface, but for different concepts, different functions are involved in the $W$ of the equation $W=W_0$. I find that, in the empirical determination of the adopted estimate $W_0$, the permanent tide is treated inconsistently. However, the consistent estimate from the same data rounds off to the same value. I discuss the tidal conventions and formulas for the International Height Reference Frame (IHRF) and the realization of the IHRS. I propose a simplified definition of IHRF geopotential numbers that would make it possible to transform between the IHRF and zero-tide geopotential numbers using a simple datum-difference surface. Such a transformation would not be adequate if rigorous mean-tide formulas were imposed. The IHRF should adopt a conventional (best) estimate of the permanent tide-generating potential, such as that which is contained in the International Earth Rotation and Reference Systems Service (IERS) Conventions and use it as a basis for other conventional formulas. The tide-free coordinates of the International Terrestrial Reference Frame (ITRF), and tide-free Global Geopotential Models (GGMs) are central in the modelling of geopotential for the purposes of the IHRF. I present a set of correction formulas that can be used to move to the zero-tide model before, during, or after the processing, and finally to the mean-tide IHRF. To reduce the confusion around the multitude of tidal concepts, I propose that modelling should primarily be done using the zero-tide concept, with the mean-tide potential as an add-on. The widespread use of the expression "systems of permanent tide" may also have contributed to the confusion, as such "systems" do not have the properties that are generally associated with other "systems" in geodesy. Hence, this paper mostly uses "concept" instead of "system" when referring to the permanent tide.

# 1 Introduction

## 1.1 Concepts

As the apparent motion of the Sun, the Moon, and the planets is concentrated above the low latitudes, the time-averages of their tide-generating potentials are not zero. At the surface of the Earth, their summed contribution is a few parts of $10^{-8}$ of the potential of the Earth. To deal with the permanent deformation that is caused to the Earth and to the gravity field by the permanent tide, two concepts (tide-free and mean-tide) are applied to the geometric shape of the Earth (which, in this context, is often called crust or topography) and three concepts (tide-free, zero-tide, mean-tide) are applied to the gravity field.

1. In the *tide-free concept* (also called *non-tidal*), the permanent deformation is eliminated from the geometric shape of the Earth. From the potential field quantities (gravity, geoid, etc.) both the tide-generating potential, and the deformation potential of the Earth (the indirect effect) are eliminated. The permanent deformation is treated using the same Love numbers ($h$, $k$, and Shida number $\ell$) as for the time-dependent tidal effects *(conventional tide-free concept)*. The *secular tide-free concept* and the secular or fluid Love numbers refer to a thought experiment: to the long-term change in the shape and gravity field of the Earth if the permanent tidal potential is completely removed, i.e., if the Sun and the Moon are removed. It is an interesting experiment for considering the dynamic flattening of the Earth, for example. To the best of my knowledge, it has never been used nor has it been suggested as a reference for geodetic quantities.

2. In the *mean-tide* concept, the permanent effect is not removed from the geometric shape of the Earth. The shape, therefore, corresponds to the long-term average under tidal forcing. The potential field includes the potential of this "average Earth", and the time-average of the tide-generating potential, although the latter is not generated by the Earth's masses.

3. The *zero-tide* concept is a "middle alternative", for the potential field quantities. The potential field is that of the "average Earth". The gravity field is generated only by the masses of the Earth (plus the centrifugal force). For the geometric shape of the Earth, the zero-tide concept is identical with the mean-tide concept.

Ekman (1989) introduced the systematic thinking about the permanent tide, terming the three different cases as "concepts" as in the above, as did Poutanen et al. (1996). Later Ekman (1996) used solely "cases". However, starting in the 1990s, there was a gradual shift in the terminology to "systems of permanent tide", within which the present author has also participated (Mäkinen and Ihde 2009). In retrospect, I think that this shift was unfortunate: The word "system" brings associations to geodetic systems like "coordinate reference systems", where formal transformations between systems are valid without consideration of the physical background of the operations. But the "systems of permanent tide" are not that kind of system. I will discuss the subject further in sections 3 and 6. For the rest of the paper until section 6, "concept" is used.

There is a lacuna in the conventional 3-point taxonomy presented above: It gives the impression that it is only the "crust" or "topography" of the Earth which could be presented either at the tide-free or at the mean-tide (=zero-tide) position. For instance, the 3-D geometric shape of the Earth is normally represented by the tide-free International Terrestrial Reference Frame (ITRF) coordinates of the topography. But, what about the coordinate representation of intangible surfaces of the potential field such as the geoid, or geoid models? A moment's reflection shows that for the potential field quantities there are two tidal concepts present: the tidal concept of the potential, and the tidal concept of the coordinate representation. The two are logically independent of each other.

There is some danger of confusion: it may be tempting to think (not in very precise terms) of the mean-tide and the tide-free coordinates as two different coordinate systems. The misleading indication "Global Geopotential Models (GGMs) are given in ITRF coordinates" may channel the users' thoughts in this direction. But, there is only one coordinate system, the system that is also used everywhere in free space, and in which GGMs are given. The instantaneous position of reference points and other objects varies periodically because of the tides. In the mean-tide concept the coordinates are given at the time-averaged position. In the (conventional) tide-free concept, the coordinates are given at a conventional off-mean position within their tidal range. However, the coordinate system is the same in both cases. This line of thought is as valid for intangible surfaces as it is for concrete objects.

Obviously, by comparing tide-free positions and mean-tide positions one can write a nonlinear coordinate transformation simulating, to some extent, their relation. But the representation of space in the "tide-free coordinate system" would bring unsurmountable problems and normal physics would fail.

In the conventional 3-point taxonomy, it was tacitly assumed that the tidally different geoids would always be represented at their mean-tide positions; see for instance Fig. 1 in Mäkinen and Ihde (2009). The question is further discussed in the example at the end of section 5.

**1.2 Historical background, current tasks**

The first time that the International Association of Geodesy (IAG) took a position on the permanent tide was at the XVII General Assembly of the IUGG (International Union of Geodesy and Geophysics) in Canberra in 1979. The tide-free concept was adopted in the IAG Resolution No. 15. This was a rapid response after Heikkinen (1979) had warned about the problems in the application of Stokes' formula that the use of mean-tide gravity, as implied by the Honkasalo (1964) correction, would cause. After this, several authors (e.g., Ekman 1979, 1981; Groten 1980, 1981) pointed out how the tide-free Earth is a problematic model for the actual Earth. At the XVIII General Assembly of the IUGG in Hamburg in 1983 the IAG then reversed its position: in its Resolution No. 9 the IAG recommended the zero-tide concept for the potential field quantities, and in its Resolution No. 16 the IAG recommended the mean-tide concept for the shape of the Earth.

The tide-free quantities that are currently in use were not a response to the IAG Resolution No. 15 of 1979, in that they were mostly born unintentionally, rather than by weighing alternatives between different tidal concepts. With respect to the early tidal corrections to gravity and levelling ("luni-solar corrections"), the new quantities were almost inevitable. The correction was made using the total tide-generating potential from (often simplified) ephemerides. It would have required a special effort to contemplate the permanent component and to care about it, never mind restore it. Later, when corrections to geodetic quantities were made using tidal spectroscopy, the method for many quantities usually was (and still is) to make, at the first step, a correction using the total tide-generating potential and then refine it for the most important waves. This is the method that has been applied in the International Earth Rotation and Reference Systems Service (IERS) Conventions, both for the geopotential and for the station positions, starting with McCarthy (1992). It was then very easy for the code-writers to overlook the fact that at the first step they also removed part of the Earth's presumed response to the permanent tide-generating potential. When Poutanen et al. (1996) pointed out that the ITRF coordinates are tide-free, the IERS Standards (McCarthy 1992) were still unambiguously prescribing mean-tide (=zero-tide) coordinates.

All three tidal concepts are currently used for referencing geodetic quantities. ITRF coordinates are tide-free (Poutanen et al. 1996). Regional and national 3-D reference frames, such as the ETRFxx (realisations of the ETRS89) derive from the ITRF and are tide-free. Their great practical importance implies that tide-free 3-D coordinates will stay with us for a long time. GGMs are provided either tide-free or zero-tide or in both versions. Legacy national height systems from levelling are either tide-free (i.e., tide-free crust over tide-free geoid) or mean-tide (mean-tide crust over mean-tide geoid). National height systems that have been adopted since 2005 are zero-tide (mean-tide crust over zero-tide geoid), as is the regional height reference frame EVRF2007 (Sacher et al. 2009). The adoption of the mean-tide concept for IHRS is now encouraging others to follow suit: the EVRF2019 update of EVRF2007 is provided in both zero-tide and mean-tide versions (Sacher and Liebsch 2020). The International Gravity Standardization Net 1971 (IGSN71) is mean-tide (Morelli et al. 1974) but all modern gravity reference values since the 1980s are provided in zero-tide (Boedecker 1988). For more history and detail, see e.g., Ekman (1989, 1996) and, especially for height systems, Mäkinen and Ihde (2009).

At the XXVI General Assembly of the IUGG in Prague, Czech Republic, in 2015 the IAG adopted the mean-tide concept for the IHRS. In its Resolution No.1, Definition and Realization of an International Height Reference System (IHRS), the IAG resolves (quoting from Drewes et al. 2016):

- the following conventions for the definition of an International Height Reference System (see note 1):
    1. the vertical reference level is an equipotential surface of the Earth gravity field with the geopotential value $W_0$ (at the geoid);
    2. parameters, observations, and data shall be related to the mean tidal system/mean crust;
    3. the unit of length is the metre and the unit of time is the second (SI);
    4. the vertical coordinates are the differences $-\Delta W_P$ between the potential $W_P$ of the Earth gravity field at the considered points P, and the geoidal potential value $W_0$; the potential difference $-\Delta W_P$ is also designated as geopotential number $C_P = W_0 - W_P$;
    5. the spatial reference of the position P for the potential $W_P = W(\mathbf{X})$ is related as coordinates $\mathbf{X}$ of the International Terrestrial Reference System;
- $W_0 = 62636853.4$ m$^2$s$^{-2}$ as realisation of the potential value of the vertical reference level for the IHRS (see note 2).

'Note 1' in the resolution above refers to Ihde et al. (2015), now available also in Ihde et al. (2017), and 'note 2' above refers to Sánchez et al. (2015), expanded to a detailed paper by Sánchez et al. (2016). Observe that item 2 means that the potential $W$ should be interpreted as the sum of the Newtonian potential of the masses of the Earth (including the potential of the permanent tidal deformation), the centrifugal potential of the Earth's rotation, and the time-average of the tide-generating potential, although the last mentioned is not always considered part of the gravity field of the Earth as it is not generated by the masses of the Earth or by its rotation.

Does the IHRS bring some new elements to the treatment of the permanent tide in height systems? After all, mean-tide height systems are not a novelty. Until recently, the overwhelming majority of national height systems were mean-tide and there has been much practice in their use, as well as in tide-free and zero-tide heights.

I do think that the IHRS now presents new questions to the way in which the permanent tide is handled, and not only in height systems. I think that this stems from three aspects:

1. Instead of reference markers realising a potential level that is derived from recent or ancient sea level observations, the datum of the IHRS is fixed by an abstract potential value number $W_0$. The relationship of the different tidal concepts to $W_0$ needs to be clarified.
2. The IHRS is global, unlike existing height systems that have maximally covered a single continent. It will be established by methods that are different from the previous mean-tide systems and the permanent tide will appear in these techniques in a different way. Conventions and corrections that are acceptable over a limited area where, primarily, height differences are treated might not be adequate in a global system.
3. The role of the permanent tide in height systems has until now been taken to mean the datum surface only: the tide-free, the zero-tide, or the mean-tide geoid. Other aspects have been treated pragmatically. But, now we are told that "parameters, observations, and data shall be related to the mean tidal system/mean crust". How rigorously should we apply this? What are the consequences if we are absolutely rigorous or alternatively if we relax the rigour a little?

In what follows, I discuss various aspects of the permanent tide in the International Height Reference Frame (IHRF), the realisation of the IHRS. As a height system does not exist in isolation from other geodetic quantities, the exposition will necessarily cover permanent-tide concepts in general.

**Notation and units.** I use the indices MT, ZT, and NT to indicate mean-tide, zero-tide, and tide-free (=non-tidal) quantities, respectively. For a quick assessment of the size of potential quantities from the perspective of say, the management of levelling networks, I occasionally use the "geopotential unit", gpu ( 1 gpu = $10\,\text{m}^2\text{s}^{-2}$ ). Thus, 1 mgpu corresponds to approximately 1 mm in height. For the same reason, formulas are presented to the precision of 0.01 mgpu, which is a usual computation precision in precise levelling.

Section 2 presents general results on the permanent tide and aims to clarify its relationship with the reference potential $W_0$. After reviewing different determinations of the time-average of the tide-generating potential, I propose to use for the IHRF the function of the IERS Conventions but with a different and more transparent normalisation. Section 2.4 discusses tide-free coordinates from the ITRF, and tide-free GGMs, which play a central role in gravity field modelling. IHRF requires zero-tide potentials as a stepping-stone to the final mean-tide potentials and mean-tide coordinates: I provide specific formulas to correct for the tide-free quantities at different phases of the modelling.

Section 3 points out that the permanent tide was treated inconsistently in the empirical estimation of $W_0$ that was the basis for the IAG adoption of the IHRS conventional $W_0$. The consistent estimate differs by $+0.0943\,\text{m}^2\text{s}^{-2}$ from the estimate preferred by Sánchez et al. (2015, 2016). However, after the rounding off to $0.1\,\text{m}^2\text{s}^{-2}$ precision, the consistent estimate agrees with the IAG conventional value.

Section 4 treats the practical and theoretical difficulties that the (minor) dependence of the permanent tide-generating potential $W_T$ on height could cause for the IHRF mean-tide geopotential numbers. In section 5, I then propose a solution: use the mean-tide geoid as a reference surface for the IHRF geopotential numbers but eliminate the height dependence of $W_T$ from them by convention. This amounts to the way the permanent tide is treated when national and regional mean-tide height systems are created using levelling networks.

## 2 General results on permanent tide

### 2.1 Basic relations

In the spectral decomposition of the tide-generating potential, only the even-degree zonal tides have non-zero time averages (Zadro and Marussi, 1973). We have

$$W_T(r,\phi) = W_{T,2}(r,\phi) + W_{T,4}(r,\phi) + \ldots = B_2\left(\frac{r}{R}\right)^2 P_2(\sin\phi) + B_4\left(\frac{r}{R}\right)^4 P_4(\sin\phi) + \ldots$$

(1)

Here,

- $W_T(r,\phi)$ is the sum of the time-averages of the tide-generating potential for Sun, Moon, and planets
- $(r,\phi)$ are the geocentric radius and latitude, respectively
- $W_{T,i}(r,\phi)$ is the spectral component of degree $i$
- $P_i(\cdot)$ is the Legendre polynomial of degree $i$
- $R$ is a scaling factor for distances
- $B_i$ is a coefficient that depends on $R$, such that for another choice of $R$ (say $R'$) the corresponding coefficient $B_i'$ fulfils $B_i'/(R')^i = B_i/R^i$

When we only retain the terms $W_{T,i}(r,\phi)$ that are at least $0.0001\,\text{m}^2\text{s}^{-2}\,(=0.01\,\text{mgpu})$ in absolute value and select $R = a$, where $a$ is the semi-major axis of the GRS80 ellipsoid, we obtain at the epoch 2000.0

$$W_T(r,\varphi) = \left[-1.94438\ \text{m}^2\text{s}^{-2}\right]\left(\frac{r}{a}\right)^2 P_2(\sin\varphi) + \left[0.00011\ \text{m}^2\text{s}^{-2}\right]\left(\frac{r}{a}\right)^4 P_4(\sin\varphi) \quad (2)$$

where the coefficients have been derived from the KSM03 tidal expansion (Kudryavtsev 2004, 2007)[1]. They agree with the digits shown with the HW95 tidal expansion (Hartmann and Wenzel 1995a, 1995b). The coefficient of the second-degree term changes by $-0.00063$ m²s⁻² per century, due to the change in the inclination of the ecliptic. The second-degree term is generated by the Moon, the Sun, and the planets – the contributions of the planets sum to one unit in the last decimal shown. The fourth-degree term is generated by the Moon; the contribution of the Sun and the planets are negligible. In tidal spectroscopy the second-degree term with the Sun and the Moon (sometimes also including the planets) is usually called $M_0S_0$. I will return to the numerical values later in this paper.

In what follows, I drop the fourth-degree term, although it has the size of the last digit (0.01 mgpu) which is traditionally carried over in e.g., precise levelling calculations. That is, however, done in order to decrease round-off errors (when the objective is 0.1 mgpu precision), which in the present case is not relevant. Thus, in the sequel I identify $W_T(r,\phi)$ with $W_{T,2}(r,\phi)$

$$W_T(r,\phi) \approx W_{T,2}(r,\phi) = B_2 \left(\frac{r}{R}\right)^2 P_2(\sin\phi). \qquad (3)$$

If $V$ is the potential of the Newtonian attraction of the masses of the Earth including the permanent tidal deformation, and $W_\Omega$ is the potential of the centrifugal acceleration of the Earth's rotation, the potential $W_{ZT}$ in the zero-tide case is

$$W_{ZT} := V + W_\Omega. \qquad (4)$$

Using the same notation, the potential $W_{MT}$ in the mean-tide case is

$$W_{MT} := V + W_\Omega + W_T. \qquad (5)$$

---

[1] Kudryavtsev (2007) also shows a permanent third-degree tide of 6×10⁻⁷ m²s⁻². It seems to be an artefact of the spectral analysis of the third-degree tides where the latter are discretely sampled over a limited time interval (personal information by e-mail from SM Kudryavtsev on November 1, 2017).

The generation of the conventional tide-free potential can be illustrated using the deformation response of a spherical non-rotating elastic and isotropic Earth. The Newtonian potential field of the model Earth deformed by the potential of Eq. (3) is $\tilde{V} = V + kV_T$ with

$$V_T = B_2 \left(\frac{R}{r}\right)^3 P_2(\sin\phi), \tag{6}$$

where $k$ is a Love number. If the tidal response of the Earth is modelled by forcing the Earth with the full tide-generating potential, including the time-independent part, the response will also include the contribution $kV_T$ with the same value of $k$ as for the time dependent tides, nominally $k = 0.3$. Removing the total response (time-dependent and time-independent parts) gives us the conventional tide-free potential $W_{NT}$

$$W_{NT} = V - kV_T + W_\Omega. \tag{7}$$

In modern geodetic practice, the time-dependent tidal response of a realistic rotating Earth is modelled using a large spectrum of spherical harmonics, frequency-dependent Love numbers, and taking into account the anelasticity of the mantle (e.g., Petit and Luzum 2010). Nevertheless, if there is a modelling step where the complete tide-generating potential (including the time average) is used, the corrected geopotential will have a tide-free second-degree zonal harmonic, just as in the simplified case above (Eqs. [6] and [7]), with the Love number $k$ used at that particular step. In addition, depending on the computation scheme, there might be a tide-free fourth-degree zonal term. This is treated in section 2.4, along with the restoration of the zero-tide values.

The conventional tide-free coordinates are generated analogously: by forcing the Earth with a tide-generating potential that also contains the time-independent part. The correction then removes the time-independent part with the same Love number $h$ and Shida number $\ell$ with which the time-dependent tide is corrected for.

In the "secular tide-free concept" a "secular" or "fluid limit" Love number of about $k_s = 0.93$ is found. The secular tide-free concept was never considered viable as a reference for geodetic quantities. It would create a reference very much different from physical reality. Compared with this, the traditional argument against it (Groten 1980, 1981; Angermann et al. 2016) "that the $k_s$ is poorly known and in practice unknowable" is insignificant.

Henceforth, I use "tide-free" without attributes as being synonymous with "conventional tide-free". This is in line with Chapter 2 ff. of IERS Conventions 2010 (Petit and Luzum 2010). Although Fig. 1.1 and Fig. 1.2 of Section 1.1 (op cit) appear to suggest that "tide-free" without attributes should point to secular tide-free quantities, it would be quite impractical to reserve the concise expression "tide-free" to the secular concept that is never used in geodetic referencing, and to always have to use the long "conventional tide-free" for the concept that is actually used.

Taking the zero-tide potential $W_{ZT}$ as reference we have the difference of the tide-free potential $W_{NT}$ and of the mean-tide potential $W_{MT}$ relative to it

$$W_{NT} - W_{ZT} = -kV_T \qquad (8a)$$

$$W_{MT} - W_{ZT} = W_T. \qquad (8b)$$

Equation (8a) and Eq. (8b) may appear rather symmetrical for $W_{NT}$ and $W_{MT}$. On the surface of a sphere we get for the differences $-kV_T$ and $W_T$ the same surface harmonic spectral component $P_2(\sin\phi)$ (with different coefficients). But, there is a fundamental distinction: the potential $-kV_T$ is generated by the (deformation of the) masses of the Earth and $W_T$ is not. Thus $-kV_T$ in Eq. (7) can be fused with the standard spherical harmonic representation of the Newtonian attraction $V$, while $W_T$ in Eq. (5) must be represented separately by Eq. (3). This point is sometimes overlooked, and the second-degree spherical surface harmonic obtained from $W_T$ is then imported into the second-degree zonal harmonic in a GGM[2]. This is erroneous: while it makes sense to talk about a tide-free $J_2$ or tide-

---

[2] The mean-tide ellipsoid of Burša (1995a, 1995b), recently revived by Angermann et al. (2016), was made in this way: fusing the spherical surface harmonic from the permanent tide into the

free $C_{20}$, there is no such thing as mean-tide $J_2$ or mean-tide $C_{20}$, except for surface spherical harmonics. For the same reason, the shorthand term "mean-tide geopotential" may easily result in misunderstandings about the character of the permanent tide-generating potential.

An example about the numerical values involved: Let us use the equatorial radius $a$ of the GRS80 ellipsoid as the scale parameter $R$ in Eq. (3) for the permanent tide-generating potential.

$$W_{T,2}(r,\phi) = B_2 \left(\frac{r}{a}\right)^2 P_2(\sin\phi). \qquad (9)$$

In the spherical-harmonic expansion of a zero-tide GGM using the same scale parameter $a$, the second-degree zonal term is

$$V_{20}^{ZT}(r,\phi) = GM \frac{1}{r}\frac{a^2}{r^2} C_{20}^{ZT} P_2(\sin\phi). \qquad (10)$$

Here $GM$ is the geocentric gravitational constant and $C_{20}^{ZT}$ is the zero-tide second-degree zonal coefficient. We can merge the geopotential and the permanent tide-generating potential in the surface spherical harmonic on a sphere. On the sphere $r = a$

$$V_{20}^{ZT}(a,\phi) + W_{T,2}(a,\phi) = GM \frac{1}{a}\left[C_{20}^{ZT} + B_2 \frac{a}{GM}\right]P_2(\sin\phi). \qquad (11)$$

But we cannot replace $C_{20}^{ZT}$ in Eq. (10) by $\left[C_{20}^{ZT} + B_2 \frac{a}{GM}\right]$ from Eq. (11) and claim that the result would represent "the second-degree zonal term in a mean tide GGM". If we nevertheless try to use such a construct for computations, we will get erroneous results. For instance, at the poles ($P_2(\sin\phi) = 1$, $r = b$) the permanent tide that we have erroneously embedded in the expansion would contribute $B_2(a^3/b^3)$ to the total potential, instead of the correct number $B_2(b^2/a^2)$ from Eq. (9). Using the value $-1.9444$ m²s⁻² for $B_2$, the error committed corresponds to +3.3 mm in height.

---

harmonic $J_2$. Subsequently, he used the modified $J_2$ as an input to calculating a Somigliana-Pizzetti level ellipsoid.

Thus, while 2-D displays and spherical formulas of the permanent-tide quantities, such as in Fig. 1 of Mäkinen and Ihde (2009) and Table 1 (op cit) can be instructive, they cannot replace rigorous calculations. If we treat global problems and want to use global models rigorously, we cannot handle the permanent tide as just a surface spherical harmonic on a sphere.

## 2.2 Geoids in different concepts of the permanent tide

It is frequently stated that the potential $W_0$ at the geoid is "free of zero-frequency tidal distortion", "independent of the tidal concept used", i.e., that the same $W_0$ is appropriate for the secular tide-free, conventional tide-free, the zero-tide, and the mean-tide geoid. This statement is sometimes (e.g., Burša 1995c) treated as a theorem that requires proof, for instance using Bruns' formula or an explicit form of the permanent tide-generating potential. However, the statement is better considered as the definition of the tidally-different geoids[3]. Suppose we take for instance the zero-tide geoid

$$W_{ZT} = V + W_\Omega = W_0 \qquad (12)$$

with a given value $W_0$ as a starting surface. We then "distort" it by adding the permanent tide-generating potential $W_T$. Which of the new equipotential surfaces

$$W_{MT} = W_{ZT} + W_T = W_1 \qquad (13)$$

with various values of the constant $W_1$ can be regarded as the tidally-distorted version of the original surface of Eq. (12)? Surely the answer is $W_1 = W_0$. The same logic applies between all of the tidal geoids. The distances between the geoids can then be calculated using Bruns' formula. The height of the mean-tide geoid above the zero-tide geoid is

$$\Delta N_{MT} = \frac{W_T}{g} \qquad (14)$$

and the height of the tide-free geoid above the zero-tide geoid is

$$\Delta N_{NT} = -\frac{kV_T}{g} \tag{15}$$

where $g$ is gravity.

---

[3] In order that the equality of the potentials be considered as a theorem requiring proof, we would need a definition of the distorted geoid that is independent of the equality of potentials. Otherwise, a proof can only amount to tautology.

When we consider a range of geoids with different tidal definitions but all with the same $W_0$, it is important to keep in mind that the potential function $W$, by which the equipotential surface is defined, is different in each case.

### 2.3 Numerical values for permanent-tide quantities

*Time average of the tide-generating potential*

Consistent formulas for all quantities that are related to the permanent tide can/should be derived from a (conventional or best) formula for the time-average of the tide-generating potential $W_T$. Formulas for $W_T$ are often presented in different normalisations. For instance, the IERS Conventions (2010) (Petit and Luzum 2010) use the Cartwright-Tayler-Edden normalisation

$$W_T(r,\phi) = B_2\left(\frac{r}{R}\right)^2 P_2(\sin\phi) = H_0\sqrt{\frac{5}{4\pi}}g_e\left(\frac{r}{R_e}\right)^2\left(\frac{3}{2}\sin^2\phi - \frac{1}{2}\right) \tag{16}$$

where $H_0 = -0.31460\,\text{m}$ is the height of the permanent tide (Cartwright and Tayler 1971; Cartwright and Edden 1973), $g_e = 9.79828685\,\text{ms}^{-2}$, $R_e = 6378136.55\,\text{m}$. To compare the formulas, I write them in the form

$$W_T(r,\phi) = A\left(\frac{r}{a}\right)^2\left(\sin^2\phi - \frac{1}{3}\right) \tag{17}$$

where $a = 6378137\,\text{m}$ is the semi-major axis of the GRS80 ellipsoid, and compare the coefficients $A$. Where can we obtain good estimates for $W_T$, i.e., for the coefficient $A$? That would be from the time-independent terms ($M_0S_0$) of the time-harmonic expansion of the tide-generating potential (Table 1).

Table 1. Time-average of the tide-generating potential from various spectral expansions, rewritten in the form of Eq. (17) and compared in terms of the coefficient $A$ at the epoch 2000.0. The rate of $A$ is -0.00095 m$^2$s$^{-2}$ per century. "Numerical" refers to spectral analysis of time series generated from numerical ephemeris, "analytical" refers to algebraic manipulations. For more details, see the text.

| # | Reference | Method | Ephemeris Lunar Solar | Number of terms | $A$ [m$^2$s$^{-2}$] |
|---|---|---|---|---|---|
| 1 | Doodson (1921) | Analytical | Brown | 378 | –2.9181 |
| 2 | Cartwright and Tayler (1971) Cartwright and Edden (1973) | Numerical | Brown/EJC Newcomb | 505 | –2.9165(2) |
| 3 | Büllesfeld (1985) | Numerical | Brown Newcomb | 656 | –2.9164 |
| 4 | Tamura (1987; 1993) | Numerical | DE118/LE62 | 1200 | –2.91656 |
| 5 | Xi (1989) | Analytical | Brown/EJC Meeus | 2933 | –2.91647 |
| 6 | Hartmann and Wenzel (1995a; b) | Numerical | DE200 | 12935 | –2.91657 |
| 7 | Roosbeek (1996) | Analytical | ELP2000-85 VSOP87 | 6499 | –2.91665 |
| 8 | Kudryavtsev (2004; 2007) | Numerical | DE/LE-406 | 28806 | –2.91657 |
| 9 | McCarthy and Petit (2004) Petit and Luzum (2010) | From (#2) | | | -2.9166(2) |

Except for items Nº 8 and Nº 9, the values in the column "$A$" in Table 1 were derived not from the original papers (column "Reference") but from Hartmann and Wenzel (1995b) and Wenzel (1996), where they have been renormalised to the same format ("HW95") as item Nº 6, and when necessary also updated with new astronomical constants. Where the original paper has less digits than those given in Table 1, I have put the extra digit in parentheses. The coefficient $H_0 = -0.31460$m in Eq. (16) enters Table 1 through both items Nº 2 and Nº 9. Item Nº 9 uses the renormalisation advice of Petit and Luzum (2010); the values are given under Eq. (16). The advice appears to ignore the original parameters of Cartwright-Tayler-Edden. Wenzel (1996) states that they have been taken into account (item Nº 2). This apparently leads to the minor difference between items Nº 2 and Nº 9, column "$A$".

Starting with item Nº 2, the differences in the coefficient *A* (the last column of Table 1) are maximally $0.0002$ m²s⁻² $=0.02$ mgpu only. In the rest of this paper, the value $-2.9166$ m²s⁻² derived from the IERS Conventions (2010) is used; it differs by less than $0.01$ mgpu from the latest estimates, those of Hartmann and Wenzel (1995a, 1995b), Roosbeek (1996) and Kudryavtsev (2004, 2007). It is proposed to adopt the $W_T(r,\phi)$ of IERS Conventions also for the IHRF Conventions. However, the Cartwright-Taylor-Edden normalisation is unwieldy and opaque for analysts outside of the tide community. Thus, for the IHRF we should adopt

$$W_T(r,\phi) = A\left(\frac{r}{a}\right)^2\left(\sin^2\phi - \frac{1}{3}\right) = A'\left(\frac{r}{a}\right)^2 P_2(\sin\phi) = A''\left(\frac{r}{a}\right)^2 \bar{P}_2(\sin\phi) \quad (18)$$

with $A = -2.9166$ m²s⁻², $A' = \frac{2}{3}A = -1.9444$ m²s⁻², $A'' = \frac{A'}{\sqrt{5}} = -0.86956$ m²s⁻².

Here $P_2(\cdot)$ is the second-degree Legendre polynomial and $\bar{P}_2(\cdot)$ is the second-degree fully normalised Legendre polynomial.

### Derived expressions

It is useful to derive from Eq. (18) expressions in ellipsoidal coordinates. The formula for $W_T$ in geodetic latitude and height $(\varphi,h)$, close to the surface of the GRS80 ellipsoid, reads as (Ihde et al. 2008)

$$\bar{W}_T(\varphi,h) = \left(1 + \frac{2h}{a}\right)\left(0.9722 - 2.8841\sin^2\varphi - 0.0195\sin^4\varphi\right) \quad [\text{m}^2\text{s}^{-2}] \quad (19)$$

where the overbar in $\bar{W}_T(\varphi,h)$ is not normalisation-related, but is used to avoid possible confusion due to change of variables compared with earlier notation. Equation (19) is valid at any height of the terrestrial topography.

The contribution of $W_T$ to the acceleration of free fall is

$$g_T(\varphi) = -30.49 + 90.95\sin^2\varphi + 0.31\sin^4\varphi \quad [\mu\text{Gal}] \quad (20)$$

The dependence of $g_T(\varphi)$ on $h$ is negligible and is not shown. Eq. (20) shows the value that should be added to zero-tide gravity in order to obtain mean-tide gravity (The IAG definition of gravity is zero-tide).

The ratio $\overline{W}_T(\varphi,0)/\gamma_0(\varphi)$, where $\gamma_0(\varphi)$ is the GRS80 normal gravity at the ellipsoid, can for instance be used to get an idea about the difference between metric zero-tide heights and metric mean-tide heights.

$$H_T(\varphi) = \overline{W}_T(\varphi,0)/\gamma_0(\varphi) = +99.40 - 295.41\sin^2\varphi - 0.42\sin^4\varphi \quad [\text{mm}] \quad (21)$$

## 2.4 Tide-free quantities

In the IHRF both coordinates and potential shall be mean-tide. Gravity field modelling, however, cannot be done with the mean-tide potential, as it contains the permanent tide-generating potential $W_T$, generated by masses outside the Earth. The potential $W_T$ can only be added at the end. It is straightforward to do the modelling in zero-tide. However, key inputs are tide-free: the published ITRF coordinates, and many GGMs. Therefore, many analysts prefer to work with tide-free quantities and reduce to the zero-tide at the end. This reduction is often done by using the generic formulas of Ekman (1989).

The purpose of this section is to recount how the zero-tide quantities can be restored before the computation. If the computation is, nevertheless, performed with tide-free quantities, formulas are provided to restore zero-tide at the end. They are specific to tide-free ITRF coordinates and to tide-free GGMs that are generated by applying the IERS Conventions.

*ITRF coordinates*

IERS Conventions of 2003 (McCarthy and Petit 2004) and 2010 (Petit and Luzum 2010) provide the formula for restoring the mean-tide position for ITRF tide-free Cartesian coordinates; it was already provided as an alternative in the 1996 Conventions (McCarthy 1996). The restoring formulas of the IERS Conventions (2010) are given with a precision of 0.1 mm (Eq. [7.14a] and Eq. [7.14b], Section 7.1.1.2). Ihde et al. (2008) calculated them with one more decimal, using Eq. (7.1a) and Eq. (7.2) (Section 7.1.1.1). The vector to be added is

$$\Delta\vec{r} = \left\{\left[-120.61 + 0.12 P_2(\sin\phi)\right] P_2(\sin\phi)\right\} \hat{r} + \left\{\left[-25.21 - 0.06 P_2(\sin\phi)\right]\sin 2\phi\right\} \hat{n} \quad [\text{mm}]$$
$$. \quad (22)$$

Here, $\hat{r}$ is the unit vector from the origin to the station, $\hat{n}$ is the unit vector at right angles to it in the northward direction, $\phi$ is the geocentric latitude, and $P_2(\cdot)$ is the second-degree Legendre polynomial.

It is useful to express Eq. (22) in ellipsoidal coordinates (GRS80). The projection $h_T(\varphi)$ of the vector $\Delta \vec{r}$ on the ellipsoidal normal is (Ihde et al. 2008)

$$h_T(\varphi) = 60.34 - 179.01 \sin^2 \varphi - 1.82 \sin^4 \varphi \quad [\text{mm}] \quad (23)$$

taken positive outwards. The projection of the vector $\Delta \vec{r}$ on the North-pointing normal of the ellipsoidal normal is

$$v_T(\varphi) = -25.13 \sin 2\varphi - 0.04 \sin 4\varphi \quad [\text{mm}]. \quad (24)$$

Thus, $v_T(\varphi)$ is the correction from the tide-free north coordinate to the mean-tide north coordinate in a local (north, east, up) coordinate system at station height. Eq. (23) and Eq. (24) are valid wherever Eq. (22) is. The correction to the geodetic latitude corresponding to Eq. (22) is

$$\Delta \varphi_T(\varphi) = -0.814 \sin 2\varphi - 0.004 \sin 4\varphi \quad [\text{mas}]. \quad (25)$$

It decreases insignificantly (in absolute value) with increasing height of the station above the ellipsoid: at 10 km the coefficient of the $\sin 2\varphi$ term is –0.813.

When the potential $W_{NT}$ or $W_{ZT}$ is to be evaluated using a geopotential model and tide-free 3-D coordinates from ITRF, firstly, the mean-tide position should be restored using Eq. (22). If the GGM is, nevertheless, evaluated at the ITRF tide-free position, the correction $\Delta W$ to the GGM + the centrifugal potential, corresponding to Eq. (22), can be calculated from Eq. (23), multiplying it by (–g) where g is gravity. We can get a good estimate of the correction by replacing g with the normal gravity at ellipsoid

$$\Delta W^{ITRF}(\varphi) \approx (-\gamma_0(\varphi)) h_T(\varphi) = -0.5901 + 1.7475 \sin^2 \varphi + 0.0273 \sin^4 \varphi \quad [\text{m}^2\text{s}^{-2}] \quad (26)$$

Eq. (26) is in error for $\Delta W^{ITRF}$ to the extent that $\gamma_0$ differs from $g$, that is, up to a couple of parts per thousand. Nevertheless, it is presented with all of the decimals, since for each individual point it can be corrected by scaling it with $g/\gamma_0$. The worst-case error in $\Delta W^{ITRF}$ without rescaling is less than 1 mgpu.

*Restoring the zero tide to GGMs and consequences for the potential*

Petit and Luzum (2010) provide a formula (6-14 in Section 6.2) for restoring the zero tide to the tide-free fully-normalised zonal coefficient $\bar{C}_{20}^{NT}$, obtained by processing the solid Earth tides with the IERS Conventions. I will look at the general formulas. The second-degree zonal term of a GGM is for fully normalised spherical harmonics

$$V_{20}(r,\phi) = GM \frac{1}{r} \bar{C}_{20} \left(\frac{r_0}{r}\right)^2 \bar{P}_2(\sin\phi) \qquad (27)$$

where GM and the scale parameter $r_0$ are specific to the model. Normally, we are free to re-scale GGMs but here $r_0$ should be the scale that was originally used in the processing of satellite gravity observations, i.e., to scale the effects of the solid Earth tide on the geopotential. If $\bar{C}_{20}$ in Eq. (27) has been provided tide-free, it means that in the model the permanent tide-generating potential was part of the forcing by zonal tides. Thus, a part of the Earth's actual contribution to zero-tide $\bar{C}_{20}$ (there is no mean-tide $\bar{C}_{20}$) was already removed together with the time-dependent part. As in Eq. (3), Eq. (6), and Eq. (8a), at the surface $r = r_0$ the presumed change $\Delta V_{20}(r,\phi)$ in $V_{20}(r,\phi)$ induced by forcing by the permanent tide-generating potential is equal to the value of the permanent tide-generating potential at this surface, multiplied by the Love number $k_{20}$. We do not need to first renormalise the expression Eq. (18) by $r_0$. Instead and equivalently we just evaluate Eq. (18) (in this case the third form) at $r = r_0$

$$\Delta V_{20}(r_0,\phi) = k_{20} A'' \left(\frac{r_0}{a}\right)^2 \bar{P}_2(\sin\phi). \qquad (28)$$

Thus, at the surface $r = r_0$ we can solve for the relation between the tide-free $\bar{C}_{20}^{NT}$ and zero-tide $\bar{C}_{20}^{ZT}$

$$V_{20}^{NT}(r_0,\phi) = V_{20}^{ZT}(r_0,\phi) - \Delta V_{20}(r_0,\phi) \qquad (29a)$$

$$GM \frac{1}{r_0} \bar{C}_{20}^{NT} \left(\frac{r_0}{r_0}\right)^2 \bar{P}_2(\sin\phi) = GM \frac{1}{r_0} \bar{C}_{20}^{ZT} \left(\frac{r_0}{r_0}\right)^2 \bar{P}_2(\sin\phi) - k_{20} A'' \left(\frac{r_0}{a}\right)^2 \bar{P}_2(\sin\phi)$$
$$\qquad (29b)$$

$$\overline{C}_{20}^{ZT} = \overline{C}_{20}^{NT} + k_{20} \frac{r_0}{GM} A'' \left(\frac{r_0}{a}\right)^2. \tag{29c}$$

From Eq. (29c) we get for the corresponding zonal components of the GGM

$$V_{20}^{ZT}(r,\phi) = V_{20}^{NT}(r,\phi) + k_{20} A'' \left(\frac{r_0}{a}\right)^2 \left(\frac{r_0}{r}\right)^3 \overline{P}_2(\sin\phi). \tag{30}$$

Obviously, the formulas would look simpler if we would also have the same scaler $r_0$ in Eq. (18). In view of the accuracy that is needed, in Eq. (29c) and Eq. (30) we can consider $(r_0/a) \approx 1$ in any case.

The IERS Conventions starting with McCarthy (1996) have the same $GM = 3.986004415 \times 10^{14}$ m$^3$s$^{-2}$, $r_0$=6378136.55 m, and $k_{20} = 0.30190$. It appears that most of the recent GGMs were calculated using these values or values that are sufficiently close to them. The correction term of Eq. (29c) agrees with the formula (6-14, Section 6.2) by Petit and Luzum (2010) within the number of significant digits that are involved in their computation.

In a tide-free GGM that is produced by observing the IERS Conventions (Petit and Luzum 2010), there is also a tide-free fourth-degree zonal coefficient $\overline{C}_{40}$. It comes from the correction to fourth-degree geopotential coefficients due to second-degree tides and has nothing to do with fourth-degree tides. From Eq. (6.6) and Eq. (6.7), p. 71 (op cit) the time-dependent tidal corrections $\Delta_t \overline{C}_{20}$ and $\Delta_t \overline{C}_{40}$ to fully normalised $\overline{C}_{20}$ and $\overline{C}_{40}$, respectively, are related by

$$\Delta_t \overline{C}_{40} = \frac{k_{20}^+}{k_{20}} \Delta_t \overline{C}_{20} \tag{31}$$

with $k_{20}^+ = -0.00089$. Taking the average in time of Eq. (31), we obtain the differences between the tide-free and zero-tide values of the coefficients

$$\overline{C}_{40}^{NT} - \overline{C}_{40}^{ZT} = \frac{k_{20}^+}{k_{20}} \left(\overline{C}_{20}^{NT} - \overline{C}_{20}^{ZT}\right). \tag{32}$$

Using Eq. (29c) and putting $(r_0/a) \approx 1$

$$\overline{C}_{40}^{ZT} = \overline{C}_{40}^{NT} + k_{20}^+ A'' \frac{r_0}{GM}. \tag{33}$$

The fourth-degree zero-tide and tide-free zonal components of the GGM are then related by

$$V_{40}^{ZT}(r,\phi) = V_{40}^{NT}(r,\phi) + k_{20}^{+} A'' \left(\frac{r_0}{r}\right)^5 \overline{P}_{40}(\sin\phi). \tag{34}$$

The coefficient of $(r_0/r)^5$ in the correction term (the second term at right) is

$$k_{20}^{+} A'' \overline{P}_{40}(\sin\phi) \approx 0.0023\left(\tfrac{35}{8}\sin^4\phi - \tfrac{15}{4}\sin^2\phi + \tfrac{3}{8}\right) \quad [\text{m}^2\text{s}^{-2}] \tag{35}$$

where the function of geocentric latitude in parenthesis ($= \overline{P}_{40}(\sin\phi)$) is maximally one in absolute value. Thus, the correction term in Eq. (34) is maximally 0.23 mgpu in absolute value. It is not clear in which tide-free GGMs the tidal correction to $\overline{C}_{40}$ according to Eq. (6.7) of Petit and Luzum (2010) was in fact applied. Therefore, the correction term of Eq. (34) is not included here in corrections to tide-free GGMs.

Now, suppose that the potential values of the GGM were evaluated using the tide-free version and we want to restore the zero-tide potential a-posteriori. Denote by $\Delta W^{GGM}$ the correction sought

$$W_{ZT}^{GGM} = W_{NT}^{GGM} + \Delta W^{GGM}. \tag{36}$$

From Eq. (30) we have

$$\Delta W^{GGM} = k_{20} A'' \left(\frac{r_0}{a}\right)^2 \left(\frac{r_0}{r}\right)^3 \overline{P}_{20}(\sin\phi). \tag{37}$$

Putting $r_0 = 6378136.55$ m and expressing $\Delta W^{GGM}$ in ellipsoidal coordinates (GRS80)

$$\Delta \overline{W}^{GGM}(\varphi, h) = k_{20}\left(1 - \frac{3h}{a}\right)\left(0.9722 - 2.8673\sin^2\varphi - 0.0690\sin^4\varphi\right) \quad [\text{m}^2\text{s}^{-2}]. \tag{38}$$

Eq. (38) after the Love number looks quite different from Eq. (19) even when neglecting the height dependence. The explanation is obvious: In Eq. (19) we have the permanent tide-generating potential itself. In Eq. (38) we have the presumed Earth response. The two expressions agree at the equator.

For older GGMs the values $k_{20}$ and $r_0$ may be different from those used in the IERS Conventions. Those values (say $k_2$, $r_1$) can be found in the documentation of the GGM and inserted into Eq. (29c). Then, Eq. (38) is not strictly valid but an inspection of Eq. (37) shows how Eq. (38) can be scaled for ($k_2$, $r_1$).

If the GGM has a tide-free $\bar{C}_{40}$ that was created according to the IERS Conventions 2010, the additional correction

$$\Delta \bar{W}_4^{GGM}(\varphi) = 0.0023\left(\tfrac{35}{8}\sin^4\varphi - \tfrac{15}{4}\sin^2\varphi + \tfrac{3}{8}\right) \quad [\text{m}^2\text{s}^{-2}] \quad (39)$$

can be deduced from Eq. (35) and summed to Eq. (38).

Except for Eq. (39), the formulas with ellipsoidal coordinates in this section were obtained by fitting their coefficients to the corresponding closed expressions, usually in Cartesian or spherical coordinates. The formulas are precise within the last decimal given. Of course, one never needs to use them: even after the computation in tide-free ITRF one can evaluate the GGM (plus the centrifugal potential) at the two positions provided by the vector of Eq. (22) and use the difference as a correction instead of Eq. (26). Similarly, when correcting a posteriori for tide-free GGM, one can simply evaluate Eq. (37) instead of Eq. (38). Expressions with ellipsoidal coordinates may provide a little more "feel" for the quantities though.

*Combine the corrections for tide-free ITRF coordinates and for tide-free GGM: a levelling analogy*

We have seen that the corrections for tide-free ITRF coordinates and for tide-free GGM are independent, both theoretically and practically. The pseudo-obligatory binding of "tide-free crust" and "tide-free potential" to a single tide-free concept originated with precise levelling and ceased to be valid when GGMs became the method with which to evaluate potential values at a large scale.

Nevertheless, it is interesting to see what happens when we combine the two corrections Eq. (26) and Eq. (38). The former corrects the potential for the tide-free ITRF coordinates, and the latter corrects for the tide-free GGM. We neglect the height dependence in Eq. (38) and use $k_{20} = 0.30190$. Then

$$\Delta W^{ITRF} + \Delta \bar{W}^{GGM} = -0.2966 + 0.8819\sin^2\varphi + 0.0065\sin^4\varphi \quad [\text{m}^2\text{s}^{-2}]. \quad (40)$$

If we work with tide-free ITRF coordinates and with a tide-free GGM but, nevertheless, skip both of the two corrections going into Eq. (40) then we have evaluated the tide-free geopotential at the tide-free coordinates. This is an analogy to tide-free geopotential numbers from levelling. They are interesting for us because many analysts who work with ITRF tide-free coordinates and tide-free geopotential verify their results by comparing them with tide-free levelling results.

If we keep the coordinate representation at mean-tide, as we normally should, the tide-free geopotential number $C_{NT}$ is (from Eq. [7])

$$C_{NT} = W_0 - (W_{ZT} - kV_T) = C_{ZT} + kV_T = C_{ZT} + \Delta W^{GGM}. \qquad (41)$$

The last form of Eq. (41) refers specifically to Eq. (37). In the levelling analogy we define

$$\tilde{C}_{NT} := C_{ZT} + \Delta W^{ITRF} + \Delta \overline{W}^{GGM} \qquad (42)$$

with $\Delta W^{ITRF} + \Delta \overline{W}^{GGM}$ from Eq. (40).

Now, a tide-free geopotential number arising from a tide-free correction to levelling is not necessarily influenced by the permanent tide-generating potential exactly in the same way as $\tilde{C}_{NT}$ is. Details will depend on the correction. It is unlikely to contain the complicated tidal response of the shape and potential of the Earth that the ITRF and GGM corrections above imply. Suppose that the levelling correction derives from an accurate computation of the gradients of the tide-generating potential (i.e., the tilt) in combination with nominal Love numbers $h_2$ and $k_2$. If we take $h_2$=0.605 and $k_2$=0.30190 (the best we can do to mimic $\Delta W^{ITRF}$ and $\Delta \overline{W}^{GGM}$), the contribution of the permanent tide to $\tilde{\tilde{C}}_{NT}$ from levelling will be

$$\tilde{\tilde{C}}_{NT} = C_{ZT} + (-h_2 + k_2)\overline{W}_T(\varphi) \qquad (43a)$$

with

$$(-h_2 + k_2)\overline{W}_T(\varphi) = -0.2947 + 0.8742 \sin^2 \varphi + 0.0059 \sin^4 \varphi \quad [\text{m}^2\text{s}^{-2}]. \qquad (43b)$$

The coefficients in Eq. (40) and Eq. (43b) differ by less than 1 mgpu. Levelling is a relative technique, such that we should be comparing differences of the geopotential numbers within Eq. (40) with differences within Eq. (43b). They, too, differ by less than 1 mgpu. Within this accuracy, tide-free precise levelling can be compared with tide-free potential modelling.

Many of the formulas in this section update and supersede the popular formulas by Ekman (1989). The context, however, is different because Ekman (1989) presented formulas for transforming between different tidal concepts. Here, the perspective is strictly that of correcting tide-free quantities to obtain zero-tide/mean-tide quantities. As to the numerical differences between the formulas, Ekman worked in a spherical approximation evaluating potentials on the surface of a sphere, and his tide-free model is a generic model. Here, there are two specific tide-free models: ITRF coordinates and tide-free GGMs generated by IERS computation schemes were addressed. Ekman (1989) was only concerned with differences and one-mm accuracy. Comparisons show that the differences of his formulas to mine are of the order of one millimetre only. Since the flattening of the Earth corresponds in proportion to 1 mgpu of the permanent tide-generating potential, one millimetre is the best that one can achieve in a spherical approximation.

## 3 Permanent-tide in the empirical estimation of $W_0$

Sánchez et al. (2016) published an estimation of the $W_0$ value that provides the best approximation of the global mean sea level (MSL). Their method amounts to averaging the potential estimates obtained from GGMs over the 3-D surface of MSL determined by satellite altimetry. They perform the calculation with three different tidal concepts and find three different results. In view of the discussion in Section 2.2 of the present paper, this is baffling. What is the reason for the outcome?

Denote by $W_{ZT}^{GGM}$ a zero-tide GGM, with the centrifugal potential included. The corresponding "mean-tide GGM" (with some terminological inaccuracy) is $W_{MT}^{GGM} = W_{ZT}^{GGM} + W_T$ where $W_T$ is the permanent tide-generating potential. The tide-free GGM is $W_{NT}^{GGM} = W_{ZT}^{GGM} - kV_T$. Now, in estimating $W_0$ each tidal version of the GGM should be averaged over a surface which approximates the geoid of the corresponding tidal concept, i.e., over a surface where in the idealised case the corresponding potential would be constant. Only then can we hope for consistent results. Denote by $S_{MT}$, $S_{ZT}$, $S_{NT}$ such approximating 3-D surfaces over the pertinent sea domain, for the mean, zero, and tide-free potential, respectively. In the idealised case (in the absence of steric, atmospheric, salinity, and ocean dynamic effects, etc.) we would expect the MSL to follow an equipotential surface of the mean-tide concept. This, after all, is the whole rationale of adopting the mean-tide concept for the IHRS. The MSL is an $S_{MT}$, while the corresponding $S_{ZT}$ and $S_{NT}$ must be computationally constructed.

Thus, the average $\hat{W}_{01}$ of $W_{MT}^{GGM} = W_{ZT}^{GGM} + W_T$ over $S_{MT}$, the 3-D representation of the MSL is a tidally consistent estimate of $W_0$. This is the mean-tide estimate of Sánchez et al. (2016). How can we construct the other surfaces? None of the surfaces $S_{MT}$, $S_{ZT}$, $S_{NT}$ is a geoid but in order to provide consistent estimates they must be spaced as the corresponding geoids. Let P be a point at $S_{MT}$. If P' and P'' are the corresponding points at $S_{ZT}$ and $S_{NT}$, respectively, the defining property of these other surfaces is $W_{MT}(P) = W_{ZT}(P') = W_{NT}(P'')$. But, since $W_{ZT}^{GGM}$ is a very good estimate of $W_{ZT}$ we have for the differences

$$W_{ZT}^{GGM}(P') - W_{MT}^{GGM}(P) = W_{ZT}(P') - W_{MT}(P)$$
$$W_{NT}^{GGM}(P'') - W_{MT}^{GGM}(P) = W_{NT}(P'') - W_{MT}(P)$$
$$\Rightarrow W_{MT}^{GGM}(P) = W_{ZT}^{GGM}(P') = W_{NT}^{GGM}(P'') . \qquad (44)$$

Thus, the GGMs with other tidal components, and their surfaces, bring nothing new, it is the same estimate repeated three times over.

What will happen if the potential models and the surfaces are paired in another way? Suppose that, for instance, we average the zero-tide potential model $W_{ZT}^{GGM} = W_{MT}^{GGM} - W_T$ over the MSL surface, as Sánchez et al. (2016) do when they calculate their "zero-tide" estimate. We will get the consistent estimate $\hat{W}_0$ plus the average of $-W_T$ over the region in question. The averaging of the tide-free potential $W_{NT}^{GGM} = W_{MT}^{GGM} - W_T - kV_T$ over the MSL produces the consistent estimate $\hat{W}_0$ plus the average of $-W_T - kV_T$. Since, in the spherical-surface approximation $V_T \approx W_T$, we see that any mis-pairing produces the consistent estimate $\hat{W}_0$ plus a bias term, which is the average of the permanent tide-generating potential $W_T$ multiplied by one of the constants $(1, -1, k, -k, 1+k, -1-k)$. The average of $W_T$ (and of $V_T$) over the ellipsoidal surface is nearly zero, hence, if we would be able to integrate over the whole globe, the biases would almost disappear.

So far, we have used the same coordinates (mean-tide) for all three geopotential models. Should we not use tide-free coordinates as well? No, we should not because using them would be inconsistent with our purpose. We want to calculate the average of the potential model over the position of the MSL. For this purpose, the position of the Mean Sea Level must not be taken at the tide-free coordinates, which represent an off-Mean position. (Warning: the argument has absolutely nothing to do with ocean tides.)

I have discussed the tidal issues around $W_0$ at a non-technical, first-principles level because I think that is where their features appear most clearly. Nevertheless, it is instructive to examine how they show up in the linearised set-up where Sánchez et al. (2016) estimate $W_0$.

Take the point P at the 3-D representation of the mean sea surface and denote by $U$ the GRS80 normal potential, by Q the point at the surface of the GRS80 ellipsoid, corresponding to P, by $W_{ZT}^{GGM}$ as before a zero-tide geopotential model, and by $W_T$ the permanent tide-generating potential. Then

$$W_T(P) + W_{ZT}^{GGM}(P) = W_T(P) + W_{ZT}^{GGM}(P) - U(P) + U(P) - U(Q) + U(Q)$$
$$= W_T + T_P - h_P \bar{\gamma}_P + U_0$$
(45)

Here, $T_P$ is the anomalous potential from $W_{ZT}^{GGM}$ and $U$ at P, $h_P$ is the ellipsoidal height of P, $\bar{\gamma}_P$ is the mean normal gravity over $\overline{PQ}$, and $U_0$ is the normal potential at the GRS80 ellipsoid. $W_0$ is estimated as

$$\hat{W}_0 = U_0 + \text{avg}_S(W_T + T_P - h_P \bar{\gamma}_P)$$
(46)

where I have denoted by $\mathrm{avg}_S(\cdot)$ the average taken over the region $S$ of the ellipsoidal surface corresponding to the sea region in question. Sánchez et al. (2016) take a weighted average, with $1/\overline{\gamma}_P^2$ as weights (Eq. [12], op cit); my notation and the discussion below cover both a weighted and a non-weighted average.

I have argued above that the potential and the surface must be matched. Thus, when the point P is on the MSL, we must include the permanent tide-generating potential $W_T$ in the averaging as I just did above in Eqs. (45) and (46). Then we get the mean-tide estimate of Sánchez et al. (2016).

If we exclude $W_T$ from the averaging, i.e., average $W_{ZT}^{GGM}$ only, we must evaluate it at the surface $S_{ZT}$. With P' at $S_{ZT}$, we have $W_{ZT}^{GGM}(\mathrm{P'}) = W_T(\mathrm{P}) + W_{ZT}^{GGM}(\mathrm{P})$ in Eq. (45). The tide-free function $W_{NT}^{GGM} = W_{ZT}^{GGM} - kV_T$ is evaluated at the point P" at the surface $S_{NT}$ with $W_{ZT}^{GGM}(\mathrm{P"}) - kV_T(\mathrm{P"}) = W_T(\mathrm{P}) + W_{ZT}^{GGM}(\mathrm{P})$. Thus, even in the linearised setup we get the same averaging three times over.

The mean-tide estimate by Sánchez et al. (2016) is the correct estimate out of their three estimates. I will now look at the other two. In their zero-tide alternative, they average $W_{ZT}^{GGM}$ over the MSL. The resulting estimate is $\hat{W}_{01}$

$$\begin{aligned}\hat{W}_{01} &= U_0 + \mathrm{avg}_S(T_P - h_P \overline{\gamma}_P) = U_0 + \mathrm{avg}_S(T_P + W_T - h_P \overline{\gamma}_P - W_T) \\ &= U_0 + \mathrm{avg}_S(T_P + W_T - h_P \overline{\gamma}_P) + \mathrm{avg}_S(-W_T) = \hat{W}_0 - \mathrm{avg}_S(W_T)\end{aligned}. \quad (47)$$

Comparison with Eq. (46) shows that $\hat{W}_{01}$ is biased by the amount of $-\mathrm{avg}_S(W_T)$. If we would be able to integrate over the whole ellipsoid, the bias (whether from a weighted or a non-weighted average) would nearly disappear. (It would not be exactly zero.) As things are, the Legendre polynomial $P_2(\sin(\phi))$ in $W_T$ means that the bias will decisively depend on the latitude range and ocean mask used.

In their tide-free alternative, Sánchez et al. (2015, 2016) average the tide-free potential model $W_{ZT}^{GGM} - kV_T$ over the MSL which they shift to its tide-free position. They obtain the tide-free position by replacing $h_P$ in (25) by $h_P + \Delta h_T$ where $\Delta h_T$ is the projection on the ellipsoidal normal of the vector from a 3-D mean-tide to ITRF tide-free coordinates. Here $\Delta h_T = -h_T$ where $h_T$ (Ihde et al. 2008, Eq. [5-7]) is the same formula as Eq. (23)[4]. We get the estimate

$$\begin{aligned}
\hat{W}_{02} &= U_0 + \text{avg}_S(T_P - k_T V - (h_P + \Delta h_T)\bar{\gamma}_P) \\
&= U_0 + \text{avg}_S(T_P + W_T - h_P \bar{\gamma}_P - kV - W_T - \Delta h_T \bar{\gamma}_P) \\
&= U_0 + \text{avg}_S(T_P + W_T - h_P \bar{\gamma}_P) + \text{avg}_S(-kV - W_T - \Delta h_T \bar{\gamma}_P) \\
&= \hat{W}_0 + \text{avg}_S(-kV - W_T - \Delta h_T \bar{\gamma}_P)
\end{aligned} \qquad (48)$$

To first order

$$\Delta h_T \bar{\gamma}_P = h \frac{-W_T}{g} \bar{\gamma}_P = -hW_T \qquad (49)$$

where $h \approx 0.61$ is the nominal second-degree Love number used in the IERS Conventions, and $g$ is gravity. Further $V_T \approx W_T$ on the ellipsoid (within flattening accuracy) and we get the approximate bias

$$\hat{W}_{02} = \hat{W}_0 + \text{avg}_S(-kV - W_T - \Delta h_T \bar{\gamma}_P) = \hat{W}_0 + \text{avg}_S(-kW_T - W_T + hW_T). \qquad (50)$$

In overview

$$\begin{aligned}
\hat{W}_{01} &= \hat{W}_0 - \text{avg}_S(W_T) \\
\hat{W}_{02} &= \hat{W}_0 - (1 + k - h)\text{avg}_S(W_T)
\end{aligned} \qquad (51)$$

The ratio of the biases is

$$\frac{\hat{W}_{02} - \hat{W}_0}{\hat{W}_{01} - \hat{W}_0} = \frac{-(1+k-h)\text{avg}_S(W_T)}{-\text{avg}_S(W_T)} = 1 + k - h \approx 0.69 \qquad (52)$$

given that $k \approx 0.30$. The numerical values by Sánchez et al. (2015, 2016) produce the corresponding ratio as –0.0665/(–0.0943)=0.705, which is close enough to Eq. (52). The negative signs of the biases of their "tide-free" and "zero-tide" estimates are obviously a consequence of the latitudes where $W_T$ is positive (up to ±35 degrees) dominating their estimates. They state that the discrepancies between their different estimates increase when the latitudinal coverage decreases, and the explanation would be the same.

---

[4] Sánchez et al. (2016, Eq. [22]) give for $\Delta h_T$ a formula which is actually that of $h_T$ but it seems that in their calculations the correct sign was used.

Sánchez et al. (2016) considered their "zero-tide" estimate to be their best estimate, $\hat{W}_{01} = 62\,636\,853.353$ m$^2$s$^{-2}$ (53)

in the notation of this paper. They did not provide an uncertainty but noted that the formal error of their estimate is about $0.02$ m$^2$s$^{-2}$. They could not explain the discrepancies between their zero-tide estimate (Eq. [53]), their mean-tide estimate

$\hat{W}_0 = 62\,636\,853.353 + 0.0943 = 62\,636\,853.447$ m$^2$s$^{-2}$ (54)

and their tide-free-estimate

$\hat{W}_{02} = 62\,636\,853.353 - 0.0278 = 62\,636\,853.325$ m$^2$s$^{-2}$ . (55)

In view of this, they rounded their zero-tide estimate (Eq. [53]) to one decimal place after the decimal point $\hat{W}_{01} = 62\,636\,853.4$ m$^2$s$^{-2}$ They recommended it to the IAG which adopted it in the resolution quoted in the Introduction to the present paper. I have argued above that only their "mean-tide" estimate (Eq. [54]) treats the permanent tide consistently. When it is rounded to one decimal place it agrees with the adopted $W_0$ .

It seems to me that the problem with the approach by Sánchez et al. (2015, 2016) is that they try to use the systems of permanent tide as if they were, well, systems. That is, they perform formal transformations between them as one would do between, say, coordinate reference systems. But tidal systems are not like that. One must always keep in mind the physical significance of the operations. I believe this case study makes for a good tutorial about what tidal concepts/systems are and what they are not. The subject is further treated in the Discussion.

# 4 Mean-tide heights in a rigorous definition?

Using the notation of section 2, the mean-tide geopotential number $C_{MT}(\mathbf{X})$ is defined by

$$C_{MT}(\mathbf{X}) = W_0 - W_{MT}(\mathbf{X}) = W_0 - [W_{ZT}(\mathbf{X}) + W_T(\mathbf{X})] = C_{ZT}(\mathbf{X}) - W_T(\mathbf{X}) \tag{56}$$

where $C_{ZT}(\mathbf{X})$ is the corresponding zero-tide geopotential number. Here $\mathbf{X}$ is any 3-D coordinate triple, say, 3-D Cartesian or 3-D ellipsoidal coordinates.

From Eq. (56), it might appear that the mean-tide geopotential numbers are obtained from the zero-tide geopotential numbers by a simple datum-surface transformation, by subtracting $W_T(\mathbf{X})$. However, $W_T(\mathbf{X})$ depends not only on latitude but also on height (see Eq. [19]). The height dependence is small, but it is there. Most of this section is dedicated to pondering what we should do about it. Existing mean-tide height systems never had to worry about this. They were constructed using precise levelling, and then the issue does not show up. The issue shows up now because IHRS is defined in 3-D space. Then, we cannot avoid taking a position on the 3-D $W_T(\mathbf{X})$.

## 4.1 Technique-related issues

The first question is whether the dependence of $W_T(\mathbf{X})$ on the height $h$ in Eq. (19) will naturally show up in some of our observations or data. Then, it would be more difficult to dismiss the height dependence in our conventions and practice.

*Global geopotential models (GGMs)* do not include the time average of the tide-generating potential, let alone its height dependence. Both are parts of the total tide-generating potential in modelling observations in satellite gravity, but they do not show up in the end-product, the GGM. The formula of Eq. (18) can be added as an extra member to a GGM (but it cannot be merged with its second-degree zonal term). The augmented GGM would then include the dependence on $h$. In the conventional proposal for IHRF geopotential numbers (section 5), if Eq. (18) is used for the IHRF stations, it would need to be evaluated at the ellipsoid, not at the observation point.

The possibility of determining potential differences using the redshift effect of the *frequency of clocks* (Bjerhammar 1975; 1985; Vermeer 1983) is progressing rapidly (e.g., Wu and Müller 2020). The frequency stability of the best clocks is now around $1 \times 10^{-18}$ (McGrew et al. 2018; Oelker et al. 2019). A frequency shift of $1 \times 10^{-18}$ corresponds to a potential difference of 0.09 $m^2s^{-2}$, which is approximately 9 mm. For geodesy, high accuracy in frequency comparisons at a distance are needed (Lisdat et al. 2016). The clocks sense the total potential including the permanent tide-generating potential, but only potential differences are accessible through clock comparisons. The accuracy needed for clock pairs situated at different elevations on topography, to detect the effect of the elevation $h$ on the permanent tide-generating potential (Eq. [19]), does not seem attainable in the near future.

In *precise levelling*, an observed height difference $\Delta h_{1,2}$ between two bench marks (1 and 2) is converted to a geopotential difference $\Delta C_{1,2}$ by multiplying it with the average gravity on the interval. In practice this is usually the average gravity of the two bench marks

$$\Delta C_{1,2} = \tfrac{1}{2}(g_1 + g_2)\Delta h_{1,2}. \tag{57}$$

Now, in a rigorous mean-tide height system, the gravity $g$ in Eq. (57) shall obviously represent the gradient of the total potential field, i.e., include $g_T(\varphi)$, the contribution of $W_T$ to gravity (Eq. 20). Note that this theoretical issue is separate and independent from the tidal correction to $\Delta h_{1,2}$, i.e., from what is the reference surface of $\Delta C_{1,2}$. This can be understood by the thought experiment of levelling straight up or down along the plumb line, where the tidal correction to $\Delta h_{1,2}$ does not enter. Using mean tidal gravity in Eq. (57) produces the rigorous dependence of the $W_T$ component in $C_{MT}$ on the height $h$ (Eq. [19]). Whether or not we should do it is a different question that will be discussed in the next sections.

## 4.2 Conversion of hypothetic rigorous mean-tide geopotential numbers to metric heights

*Orthometric heights*

The orthometric height of a point P is the distance of P from the geoid, measured along the plumb line. In this definition, the tidal type of geopotential only appears through the geoid definition (tide-free, zero-tide, mean-tide) and the plumb line, not through what quantities are contained in the geopotential number that gives the potential difference between the geoid and P. The geopotential number becomes just a computational means to an end. Thus, it would seem that there are no theoretical problems involved, when we convert a mean-tide geopotential number to an orthometric height above the mean-tide geoid by dividing it with average gravity along the plumb line. If the geopotential number contains the small height-dependent part and we want to be rigorous, then we divide it by mean-tide average gravity (i.e., zero-tide gravity augmented by $g_T(\varphi)$ from Eq. [20]). If not, we divide it by zero-tide gravity and still are rigorous. If we do not care and would rather use a mix, that would also present no problem. The situation would be comparable to an error in the gravity value at the levelling benchmark. The value often comes from interpolation and might have an uncertainty larger than the error of less than 0.1 mGal that would come from using the wrong kind of gravity in the conversion. This line of thought is equally valid for Helmert heights and rigorous orthometric heights. In section 5, however, I will propose a definition of IHRF geopotential numbers that circumvents the issues just discussed.

*Normal heights*

Normal height is the height above the ellipsoid that produces the same potential difference in the normal gravity field as the geopotential number gives in the actual gravity field. The key expression here is "normal gravity field". If the conventions are rigorously mean-tide, then we must have a normal gravity field that includes the permanent tide-generating potential. The concept of normal height is not just some pretext to conveniently allow us to divide the geopotential number with average normal gravity instead of the troublesome (if done rigorously) calculation that is needed for orthometric heights. Instead, the normal height is a building block in a rigorous theory to solve for the shape of the Earth. Making mean-tide normal heights rigorous would force us to include the permanent tide-generating potential in the normal potential field, in the same way that the potential field of the centrifugal force is already included, i.e., to enlarge upon the Somigliana-Pizzetti theory. It can be done as has been demonstrated by Vermeer and Poutanen (1997). Obviously, it would mean a complete disruption of all or of a part of the current ellipsoid-based reference system. (Vermeer and Poutanen [1997] demonstrated how the enlargement can be performed without changing the geometric ellipsoid.) The level ellipsoid is then an equipotential surface of the attraction of the Earth's masses + centrifugal force + permanent tide. Observed and normal gravity include the contribution of the permanent tide-generating potential which is eliminated from gravity anomalies, just as the centrifugal force already is. Stokes' formula would be valid. Such a model might be worthwhile in connection with a total overhaul of the ellipsoidal reference, if ever undertaken. Note that such an overhaul would not in any way interfere with the ITRS or ITRF.

In section 5, I propose a definition of IHRF geopotential numbers that excludes the height dependence of the permanent tide-generating potential.

# 5 A conventional definition for geopotential numbers in IHRF

Recall the generic relation (Eq. [56]) between mean-tide and zero-tide geopotential numbers

$$C_{MT}(\text{P}) = C_{ZT}(\text{P}) - W_T(\text{P}) \tag{58}$$

where P is the field point. The dependence of $W_T(\text{P})$ on the height of P is minuscule (Eq. [19]) but troublesome both practically and theoretically. We could just ignore it by implicit common consent. However, that would create an unclear situation about the exact definition of IHRF geopotential numbers. From section 4, it appears that the best method is to eliminate the height dependence by a convention, by replacing $W_T$ at the field point P by $W_T$ at the foot point Q at the mean-tide geoid, projected along the plumb line

$$C_{MT}^{IHRF}(\text{P}) := C_{ZT}(\text{P}) - W_T(\text{Q}) .\qquad(59)$$

The definition of Eq. (59) corresponds to the intuitive idea that most geodesists have had all along about the IHRF, i.e., that the difference between mean-tide heights and zero-tide heights should be in the datum surface only. The datum surface where $C_{MT}^{IHRF} = 0$ is the mean-tide geoid instead of the zero-tide geoid, but the potential differences on the plumb line through Q are measured the same way for both zero-tide geopotential numbers and IHRF geopotential numbers. Eq. (59) is in fact the definition of mean-tide geopotential numbers that is used in national and regional levelling networks.

Since the distance between the geoid and ellipsoid is maximally around 100 m, in practice $W_{T0} = W_T(\text{Q})$ in Eq. (59) can always be calculated from $W_T$ at the ellipsoid (cf. Eq. [19])

$$W_{T0} \approx \overline{W}_T(\varphi,0) = 0.9722 - 2.8841\sin^2\varphi - 0.0195\sin^4\varphi \quad [\text{m}^2\text{s}^{-2}]\qquad(60)$$

with an error of, maximally, one unit in the last decimal given or 0.01 mgpu. Obviously, one can also evaluate Eq. (18) at the ellipsoid.

The definition of Eq. (59) means that the IHRF geopotential number will differ (very slightly) from the "natural" or "generic" mean-tide geopotential number of Eq. (58) with the same datum at the same location. The difference on the Earth's topography will always be less than 0.3 mgpu.

The treatment of the permanent tide in $C_{MT}^{IHRF}$ is then the same as in familiar mean-tide geopotential numbers from levelling networks. The orthometric height $H_{ORT}^{IHRF}$ above the mean-tide geoid is rigorously defined and is computed from $C_{MT}^{IHRF}$ using zero-tide gravity in the standard way, either rigorously or as Helmert heights. The normal height $H_{NORM}^{IHRF}$ is formally defined in the standard way, as the height above the ellipsoid that produces the same potential difference in the GRS80 normal gravity field as $C_{MT}^{IHRF}$ represents in the actual gravity field. Mean-tide normal heights of this kind do not have a place in the Molodensky theory (see Section 4.2). This does not cause practical problems.

According to conventional wisdom, neither $H_{ORT}^{IHRF}$ nor $H_{NORM}^{IHRF}$ should be used to calculate free air gravity anomalies if the anomalies are going to be input to Stokes' formula: If the zero-tide geoid is our reference, the mean-tide geoid is at $W_T/g$. The geoid calculated using Stokes' formula and global free-air anomalies that use mean-tide heights would be at the height $-2W_T/g$ relative to the zero-tide geoid, but not at $W_T/g$ where the mean-tide geoid is. (If we in addition use mean-tide gravity, we would be at $-4W_T/g$). This argumentation with respect to Stokes' formula that was decisive at the IAG 1979 Assembly has lost its weight as the relevant wavelengths in modern geoid computations are taken from GGMs. Sánchez and Sideris (2017) show that the residual effects from disparate national height systems that are implicit in the gravity anomalies are negligible. IHRF mean-tide metric heights embedded in gravity anomalies will be harmless in the same way. Thus, there is no reason to discourage the use of the IHRF metric heights in producing gravity anomalies. On the contrary, the IHRF heights could also provide a long-overdue unification in this respect.

Eq. (21) can be used to get an idea about the differences of IHRF metric heights and corresponding zero-tide metric heights.

*Dissemination issues*

By the IHRS definition, the IHRF geopotential $C_{MT}^{IHRF}(\mathbf{X})$ numbers shall be related to mean-tide coordinates $\mathbf{X}$. Section 2.3 provided formulas on how to deal with the tide-free ITRF coordinates in geopotential modelling. The issue will emerge again when IHRF heights are disseminated. The users only have immediate access to tide-free ITRF coordinates $\mathbf{X}'$. It might be too awkward and error-prone to have them perform their own conversion from $\mathbf{X}'$ to $\mathbf{X}$ before accessing IHRF. Instead, the IHRF models that are distributed to them should contain the conversion, i.e., be expressed as a function of $\mathbf{X}'$.

**Example.** Sánchez et al. (2021, Eq. [4]) present the results of regional geopotential modelling in the form

$$W_{ZT}(P) = U(P) + \gamma(P) \cdot \zeta + W_0 - U_0 \qquad (61)$$

where the coordinates of P must be mean-tide. Here $\zeta$ is the height anomaly from the solution of the Geodetic Boundary Value Problem using the GRS80 normal gravity field as a reference, and $\gamma(P)$ is normal gravity at P. Then

$$C_{MT}^{IHRF}(P) = W_0 - W_{ZT}(P) - W_{T0} = U_0 - U(P) - \gamma(P) \cdot \zeta - W_{T0}. \qquad (62)$$

Suppose that, firstly, we want to disseminate normal heights $H_{NORM}^{IHRF}$. Then, by the definition of $\zeta$ we have

$$H_{NORM}^{IHRF}(P) = h_{ell}(P) - \zeta - H_T(\varphi) \qquad (63)$$

where $h_{ell}(P)$ is the height of P above the GRS80 ellipsoid and $H_T(\varphi)$ is defined by Eq. (21). From ITRF coordinates one does not, however, obtain $h_{ell}(P)$ but rather the tide-free $h_{ell}^{ITRF}(P)$, i.e.

$$h_{ell}^{ITRF}(P) = h_{ell}(P) - h_T(\varphi) \qquad (64)$$

where $h_T(\varphi)$ is given by Eq. (23). Thus,

$$H_{NORM}^{IHRF}(P) = h_{ell}^{ITRF}(P) + h_T(\varphi) - \zeta - H_T(\varphi). \qquad (65)$$

For a gridded/interpolated $\zeta = \zeta(\varphi, \lambda)$ outside the original computation points Eq. (65) expresses a height-anomaly/quasigeoid model

$$H_{NORM}^{IHRF}(P) = h_{ell}^{ITRF}(P) - S(\varphi, \lambda) \qquad (66)$$

where

$$S(\varphi, \lambda) = \zeta(\varphi, \lambda) - h_T(\varphi) + H_T(\varphi). \qquad (67)$$

In the model of Eq. (66) the potential is mean-tide and at first sight the coordinate representation appears to be tide-free. Is it therefore in conflict with the IHRS definition which prescribes that "parameters, observations, and data shall be related to the mean tidal system/mean crust"? No. The surface $S(\varphi, \lambda)$ incorporates the shift from the tide-free ellipsoidal height $h_{ell}^{ITRF}(P)$ to the mean-tide ellipsoidal height $h_{ell}(P)$. The minor difference between the tide-free ITRF latitude and the corresponding mean-tide latitude can be ignored in this connection (cf. Eq. [25]).

Similarly, it is customary to represent national or regional geoid models based on zero-tide/mean-tide potential in tide-free coordinates derived from the ITRF. But then the model surface invariably contains (either implicitly or explicitly) a coordinate shift for the height component, from tide-free ellipsoidal height to mean-tide ellipsoidal height. Thus, the geoid model is given at the mean-tide position. Otherwise it would not be consistent with, say, underlying precise levelling in the zero-tide/mean-tide concept. However, when the geoid model is aligned to precise levelling in the tide-free concept, then the model surface does not contain a shift to mean-tide ellipsoidal height and the model is effectively given at the tide-free position.

The geopotential number $C_{MT}^{IHRF}(\text{P})$ can be found pointwise from Eq. (66) by multiplying both sides by the mean value of the normal gravity $\gamma(h)$ over the interval $h \in [0, H_{NORM}^{IHRF}(\text{P})]$.

## 6 Summary and discussion

I have written the previous sections while consistently using "concepts" instead of "systems" of permanent tide, as I consider the term "systems" to be misleading. In this section, I will discuss the current practices where the use of "systems" is firmly embedded. Because of that I will also be using "systems".

In section 3, I demonstrated the tidal inconsistencies in the empirical estimation of $W_0$, which was the basis for the IAG adoption of the IHRS conventional $W_0$. My point is not "what should be the first digit behind the decimal point in the $W_0$". My point is that the errors that were made in the estimation of $W_0$ were a consequence of the current "systems" approach to permanent tide.

The standard way of operating with these systems, which by now is deeply ingrained in the minds of geodesists', is like this: (1) Learn the *mantra* given by the three bullet points in the Introduction (replacing "concept" by "system"); (2)

Locate where your quantity is in the taxonomy of the three entities of the permanent tide; (3) This provides an identifier for your quantity in the same way that coordinate reference systems have identifiers in transformation libraries; (4) Transform all of your quantities to the same tidal system using the identifiers thus established and formulas from the literature; (5) Now you are alright.

Except that you are not alright, as Chapter 3 shows. The problem comes from implicitly assuming that the three systems of permanent tide are equally valid for any purpose and that it is sufficient to consider formal operations between them. Now, of course there is nothing wrong with the geodesists' wish to perform operations between systems without going deep into the meaning of those systems. On the contrary, that is the very utility of systems. We do not want a situation where you must know the history and deep properties of, say, ED50 and ETRF2000 in order to be able to transform coordinates between them. But the "systems" of the permanent tide are too complicated for this simplistic mode of operating. Instead, the physical significance of the operations must always be understood.

This obviously raises the questions: if the systems of permanent tide do not help us to automatise the work and we always must consider the physics of the situation, why do we need the systems? Why not just consider the physics? Using the names tide-free, zero-tide, and mean-tide give us a convenient shorthand with which to describe quantities, but in what sense, if any, do the quantities with the same attribute form a system? Would it not be better to go back to the original terminology of Ekman (1989, 1996): "concepts" or "cases"?

One can indeed argue that the confusion might be less today if we had stayed with "concepts" or "cases". The word "systems" prompted the creation of geodetic quantities in all the "system" varieties as it was taken to imply that all such quantities are needed or useful, or they at least make sense. But some of those quantities were inconsistent, such as the "mean-tide ellipsoid" of Burša (1995a, 1995b) as explained in section 2.1. Others are of questionable physical validity, for instance, many tide-free quantities related to the physical properties of the Earth (see the discussion by Ekman [1981; 1996]).

More importantly, the illusory clarity of the "systems" removed the urgency from harmonising the treatment of the permanent tide in geodesy. After the initial embarrassment at the accidental adoption of the tide-free quantities, it was possible to take the standpoint of: "Well, the situation is not ideal, but if there is a system for these things, it is OK".

Would it help to return to "concepts" now? I do not think so. Just changing names would not make a difference anymore. We would continue to operate with "concepts" in the same way as we now operate with "systems".

Does all of this matter? After all, the error in the case discussed was the equivalent of +10 mm of sea level only. However, one might point out that this was the fortuitous result of much larger biases averaging out in the integration. The bias at the equator is +99 mm (Eq. [21]). The bias at the North Pole would be −196 mm. Further, obviously, the pitfalls in the current paradigm of treating the permanent tide are not limited to the problem of estimating $W_0$, or more generally to questions concerning the MSL. Is a paradigm that facilitates (one is tempted to write "invites") such confusion in the treatment of the permanent tide sustainable in the long run? What could be done to remedy this?

The "systems" and the quantities already produced in them are not going away. But, below, I aim to sketch one possible line of development on how to use them.
1. There are too many systems in use. The zero-tide and the mean-tide are nature's systems. The tide-free is a human system: an assemblage of mostly non-connected missteps, and of decisions with unintended consequences. In the long run, discontinue this system as a data environment.
2. In the short run, work with zero-tide quantities. As long as the normal gravity field does not contain the permanent tide (see section 4), the zero-tide system is the only set of quantities that both describes the actual Earth and facilitates the solution of the geodetic boundary value problem. To emphasise this, it might be useful to re-introduce zero-tide coordinates as a synonym for mean-tide coordinates.

3. If one cannot work in the zero-tide system, it may be useful to at least think in it, as a reality checkpoint.
4. From this perspective, the tide-free quantities are just biased quantities, but they must be managed. There are tide-free coordinates and tide-free potential models. Section 2.4 contains exact formulas to correct for the two most common biases: in the ITRF coordinates and in the tide-free potential models in line with IERS Conventions.
5. The mean-tide system for the potential is an add-on after everything else has been done. Formulas for this can be found in section 2.3, and for the IHRF geopotential numbers in section 4.
6. Maintain realism. Whenever possible, talk about tide-free/zero-tide/mean-tide potential, tide-free/zero-tide=mean-tide coordinates, etc., rather than about potential/coordinates/etc., in the tide-free/zero-tide/mean-tide/system. The increased level of abstraction seldom illuminates anything and may imply a system that, possibly, is not even there. Almost always, two tidal systems are present and are independent: the coordinates and the potential. That is not scary. Specify both.

This physically realistic framework would be easier to explain to non-specialists than the present setup. Currently, most non-specialists seem to regard the tidal systems as an esoteric subject where they do not dare to venture, a subject that needs special erudition. On the other hand, any geodesist understands immediately the two fundamental features of tide-free quantities when they are explained in basic physics language: (1) ITRF coordinates of markers are not given at the time-average of their tidal variation in position, but off-average, and (2) many geodesists prefer to model the Earth with a part of the Earth's gravity field missing.

But, when they are told that these two facts are a part of a theoretical framework called "systems of permanent tide for geodetic quantities" and of its particular branch "conventional tide-free quantities", and somehow they must go together, many of them get confused and decide that the subject is not for them. Thus, in many cases the "systems of permanent tide" mystify things, instead of clarifying them. I believe that the protocol described in sections 1–6 would have an empowering effect on geodesists in general.

The supplementary material contains a compendium of legacy formulas related to the permanent tide.

**Author Contribution.** The study was in its entirety devised and written by JM.

**Data availability.** Only data in the published references were used.

## Acknowledgment

The paper profited from discussions in the GGOS Focus Area "Unified Height System" and its working group 0.1.2: Strategy for the realisation of the IHRS. Comments by anonymous reviewers and by the editors improved the presentation considerably.

# The permanent tide and the International Height Reference Frame IHRF

Jaakko Mäkinen

**Supplementary material: Legacy formulas for the permanent tide**

I review legacy formulas for the permanent tide and for related quantities. Many of them are either still in use or have been applied in geodetic products that are still in use. Where appropriate, I compare them with more modern formulas from this paper. The general observation is that the legacy formulas were quite adequate and there is usually no need to update the quantities produced with them unless the purpose is to change the tidal system as well.

**The Honkasalo correction.** Honkasalo (1964) pointed out that the time average $c_{ls}$ of the routine tidal correction to gravity observations was not zero. That is not the case with the contemporary zero-tide correction either, but in 1964 the corrections made gravity tide-free. He calculated the average of the standard correction based on the zenith angle of the sun and the moon. With the gravimetric factor $\delta = 1.20$ he found

$$c_{ls} = 0.037\left(1 - 3\sin^2\varphi\right)[\text{mGal}] \qquad (A\text{-}1)$$

where $\varphi$ is geodetic latitude. He proposed that $-c_{ls}$ should be added to the tidal correction or to gravity values already used. That is, he proposed to use mean-tide gravity. Thus, Honkasalo's estimate of the contribution of $W_T$ to observed gravity (or rather to the observed acceleration of free fall) is

$$g_{ls} = \frac{-0.037}{1.20}\left(1 - 3\sin^2\varphi\right)[\text{mGal}] \qquad (A\text{-}2)$$

In Fig. 1a the $g_{ls}$ from Eq. (A-2) is compared with $g_T(\varphi)$ from Eq. (20).

The correction $-c_{ls}$, usually computed with the gravimetric factor of the tidal correction actually performed (say, $\delta = 1.16$), became known as the Honkasalo correction. It should signify that mean-tide gravity was calculated after a tide-free tidal correction. However, when one reads in a data description that the Honkasalo correction/term was included/not in the tidal correction/gravity values that can mean that the gravity is mean-tide, or that it is zero-tide or that it is tide-free, such that is better to study the documentations and data in detail.

The Honkasalo correction $-c_{ls}$ was applied in the IGSN71 (Morelli et al 1974). After the IAG Resolution of 1979 the IGB Working Group 2" World Gravity Standards" issued instructions how to eliminate the correction from the IGSN71 (Uotila 1980) but that does not seem have happened frequently. Even after the IAG 1983 resolution the published IGSN71 values have mostly been considered conventional and not tampered with. That they are mean-tide is taken into account when they are compared with say, new absolute gravity measurements which are zero-tide.

**Zadro and Marussi** (1973) calculated the time-average of the tide-generating potential for the second- and fourth-degree zonal tides, using basic properties of the ephemerides of the sun and the moon. Their published value for the coefficient of the second-degree Legendre polynomial ($A'$ in Eq. [18]) is $A' = -1.94 \text{ m}^2\text{s}^{-2}$. Their results were quoted extensively by Burša (1995a, 1995b, 1995c) but do not seem to have been used in processing geodetic observations.

**Heikkinen's** (1978) program of tidal corrections calculates the ephemerides of the sun and the moon using Newcomb, Brown and Eckert-Walker-Eckert, and from them directly the total tide-generating force in three dimensions. He then used his program to calculate the average of the vertical component and of the N-S component of the tide-generating force on a meridian on the surface of the GRS67 ellipsoid over 18 years for the sun and 18.6 years for the moon. The function

$$(F_{ver})_{ave} = -15.14 + 45.62 \cos(2\varphi) + 0.07 |\sin(2\varphi)| \quad [\mu\text{Gal}] \quad (A\text{-}3)$$

approximated well the vertical average (positive outwards).

For the elevation of the mean-tide geoid above the zero geoid (obviously he was not using this terminology) Heikkinen (1978) found using spherical approximation

$$N_n = 0.148\left(\cos^2(2\varphi) - \frac{1}{3}\right) = -0.296\left(\sin^2\varphi - \frac{1}{3}\right) \quad [\text{m}] \quad (A\text{-}4)$$

Heikkinen's formulas did not however reach wide circulation until Ekman (1989) adapted them for use in his three systems of the permanent tide.

**The Standard Earth Tide Committee** was set up by the Permanent Commission of Earth Tides of the IAG following a recommendation of the IAG General Assembly in 1979. The committee's report (Rapp et al 1983) served as a basis for the resolution by the 1983 General Assembly. The committee recommended zero tide, the Cartwright-Tayler-Edden harmonic expansion with its time average for the tide-generating potential ($M_0S_0$), and the Wahr (1981) theory for time-dependent tidal effects. The last-mentioned recommendation required an amendment to the Wahr tidal correction to observed gravity because it produced tide-free gravity. The correction to add to tide-free gravity to bring it to zero-tide was

$$\overline{\delta f}_c = -4.83 + 15.73 \sin^2\psi - 1.59 \sin^4\psi \quad [\mu\text{Gal}] \quad (A\text{-}5)$$

where $\psi$ is geocentric latitude.

Eq. (A-5) appeared in the International Absolute Gravity Basestation Network (IAGBN) Standards (Boedecker 1988) where it apparently has baffled many gravimetrists over the years. It is not clear to which extent it was ever used by them. Wahr's definition of the gravimetric factor differed from that adopted by gravimetrists, and when the spectral approach became popular in applications requiring high accuracy, the $M_0S_0$ term could be dealt with from the beginning.

**Project MERIT Standards** (Melbourne et al 1983) are an early form of the IERS Standards/Conventions and were used extensively in dealing with the permanent tide before them. The Cartwright-Tayler-Edden $M_0S_0$ in the epoch 1960.0 is applied and recommendations are rather similar to later IERS Standards and Conventions.

**Ekman** (1989) adapted Heikkinen's formula for the contribution of the time-average of the tidal potential to observed gravity

$$\bar{g} = -30.4 + 91.2 \sin^2 \varphi \quad [\mu\text{Gal}] \qquad (A\text{-}6)$$

where $\varphi$ is geodetic latitude. Fig 1a shows a comparison of Eq. (A-6) with $g_T(\varphi)$ from Eq. (20).

Ekman's (1989) formulas for the geoid and crust in different tidal systems come from a single quantity, his spherical approximation for $W_T/g$ in the notation of this paper, $\bar{W}/g$ in the notation by Ekman (1989). His Eq. (1) reads

$$\frac{\bar{W}}{g} = 99 - 296 \sin^2 \varphi \quad [\text{mm}] \qquad (A\text{-}7)$$

where $\varphi$ is geodetic latitude. Ekman then obtains all his formulas for different tidal versions of the crust and geoid by multiplying $\bar{W}/g$ with various linear combinations of ones and the Love numbers $h$, $k$.

In Fig. 1b I have first compared Ekman's $\bar{W}/g$ (Eq. [A-7]) with $H_T(\varphi) = \bar{W}_T(\varphi,0)/\gamma_0(\varphi)$ from Eq. (21). The quantity plotted is $\bar{W}/g - H_T(\varphi)$ (blue dashed line).

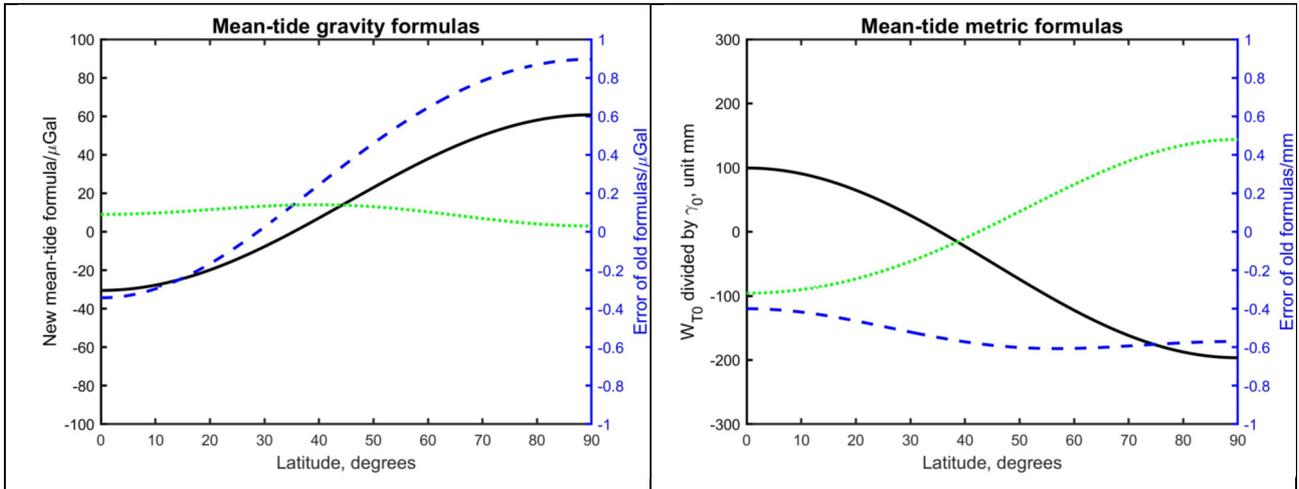

| | |
|---|---|
| **Fig. 1a**. The solid black line (left-hand scale) shows the contribution of the time-average of the tide-generating potential to the acceleration of free fall (Eq. [20]). The blue dashed line (right-hand scale) shows the difference between the same quantity deduced from the Honkasalo correction (Eq. [A-2]), and Eq. (20). The green dotted line (right-hand scale) shows the difference between Ekman's (1989) formula (Eq. [A-6]), and Eq. (20). | **Fig. 1b.** The solid black line (left-hand scale) shows $H_T(\varphi)$ from Eq. (21). The blue dashed line (right-hand scale) shows the difference $\bar{W}/g - H_T(\varphi)$ where $\bar{W}/g$ is Ekman's quantity from Eq. (A-7). The green dotted line (right-hand-scale) shows the difference $Corr_{Ekman} - Corr_{JM}$ that is explained and commented in the text. |

However, the differences between Ekman's formulas and mine do not only come from the approximation in Eq. (A-7). The spherical approximation influences different quantities in different ways. Of particular interest is the combined correction made to a geopotential number computed with the tide-free ITRF and with a tide-free GGM. It appears that Ekman's formulas are frequently used for this correction after potential modelling in the tide-free system. All Ekman's formulas are for metric heights. The best we can do with Ekman's quantity $\bar{W}/g$ (Eq. [A-7]) is to multiply it by $-(-h_2 + k_2)$ with $h_2$=0.605 and $k_2$=0.30190

$$Corr_{Ekman} = -(-h_2 + k_2)\frac{\bar{W}}{g} = -(-0.605 + 0.30190) \cdot (99 - 296 \sin^2 \varphi) \quad [\text{mm}] \tag{A-8}$$

My corresponding formula (Eq. [43b]) is a correction to the potential. I change the sign and divide by the normal gravity at the ellipsoid to make it comparable with Eq. (A-8)

$$Corr_{JM} = \frac{-\left(\Delta W^{ITRF} + \Delta \bar{W}^{GGM}\right)}{\gamma_0(\varphi)} \tag{A-9}$$

Fig. 1b shows $Corr_{Ekman} - Corr_{JM}$ in millimetres (green dotted line). Ekman's generic formula is performing very well even in this case, but its error could not be deduced from the error of his $\bar{W}/g$ (the blue dashed line).

**Additional references for the Supplementary Material**